\documentclass[lettersize,journal]{IEEEtran}
\usepackage[utf8]{inputenc}
\usepackage{color,xcolor}
\usepackage{epsfig}
\usepackage{amsmath,amsfonts,amsthm,amssymb}
\usepackage{graphics,graphicx,pgfplots,tikz}
\usepackage{float}
\usepackage{algorithm,algorithmic}
\usepackage{svg}
\usepackage{epstopdf}
\usepackage{soul,csquotes,ulem}
\usepackage[hidelinks]{hyperref}
\usepackage[font=footnotesize]{subcaption}
\usepackage[font=footnotesize]{caption}

\usepackage[subtle]{savetrees}

\begin{document}
\title{Probabilistic Positioning Via Ray Tracing With Noisy Angle of Arrival Measurements}
	\author{
		\IEEEauthorblockN{
			Vincent Corlay, Viet-Hoa Nguyen, and Nicolas Gresset
		}
		\thanks{
			The authors are with Mitsubishi Electric Research and Development Centre Europe, 35700 Rennes, France (e-mail: \{v.corlay, v.nguyen, n.gresset\}@fr.merce.mee.com).
		}
	}
	\maketitle
	

\begin{abstract}
We consider the positioning problem in non line-of-sight (NLoS) situations, where several  base stations (BS) try to locate a user equipment (UE) based on uplink angle of arrival (AoA) measurements and a digital twin of the environment.
Ray launching in a Monte Carlo manner according to the AoA statistics enables to produce a map of points for each BS. These points represent the intersections of the rays with a xy plane at a given user equipment (UE) elevation.
We propose to fit a parametric probability density function (pdf), such as a Gaussian mixture model (GMM), to each map of points. Multiplying the obtained pdfs for each BS enables to compute the position probability of the UE. This approach yields an algorithm robust to a reduced number of launched rays.
Moreover, these parametric pdfs may be fitted and stored in an offline phase such that ray tracing can be avoided in the online phase. This significantly reduces the computational complexity of the positioning method. 
\end{abstract}



\section{Introduction}

Indoor non line-of-sight (NLoS) positioning based on radio signals is a challenging problem which cannot be addresses via standard LoS-based techniques. Therefore, most solutions for this situation either rely on machine learning/fingerprinting approaches \cite{Chatelier2023}\cite{Wielandt2017}\cite{TdocSumOct} or digital twin-based techniques (often using ray tracing) \cite{Kaya2007}\cite{Kikuchi2006}\cite{Kong2006}\cite{Kong2016}\cite{Liu2014}\cite{Nguyen2023}\cite{Ryzhov2023}\cite{Voltz1994}.

\begin{figure}[t]
\vspace{-2mm}
\hspace{-4mm}
    \centering
    \includegraphics[scale=0.58]{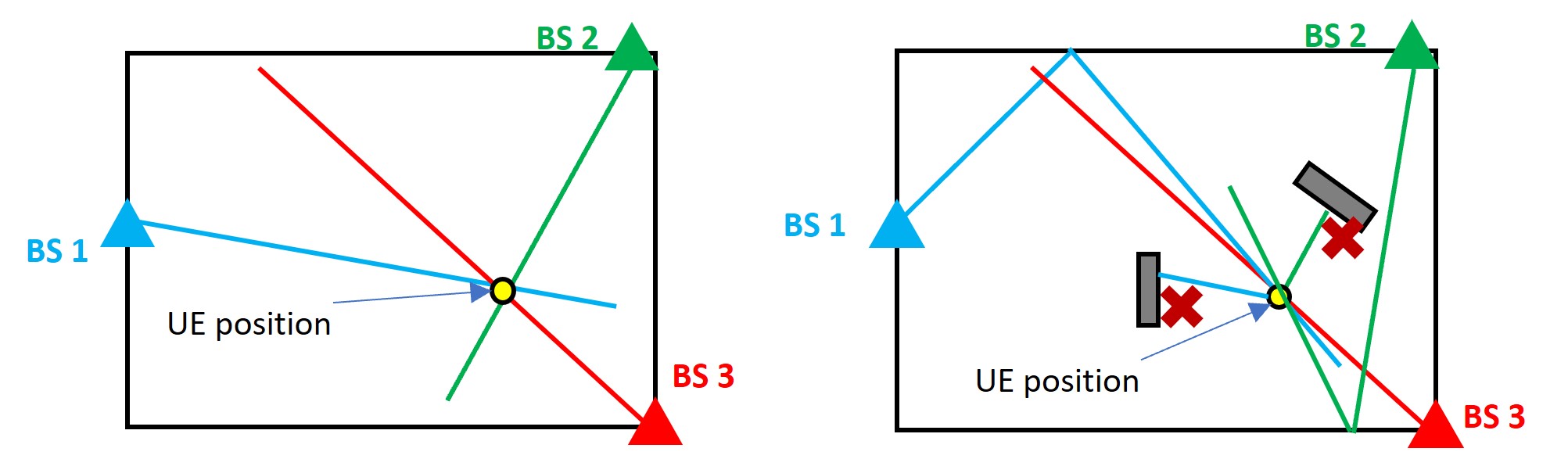}
    \caption{Left: Standard triangulation approach via  LoS-based AoA at the BS. Right: Situation where some LoS AoA are not available for several BS. In the example, two of the three LoS paths are blocked by clutters. In this case,  reverse ray tracing using a digital twin can be considered. }
    \label{fig:example_triang_reverse_ray}
\vspace{-5mm}
\end{figure}

This paper considers the following digital twin-based technique to address the NLoS uplink positioning problem. Several base stations (BS) perform angle of arrival (AoA) measurements on an uplink signal transmitted by a user equipment (UE). Then, using a digital twin of the environment, ray tracing in these AoA directions is performed. The intersection of the rays is then the estimated position of the UE. This approach,  illustrated on Figure~\ref{fig:example_triang_reverse_ray}, can be called “reverse/backward ray tracing” \cite{Kaya2007}\cite{Kong2006}\cite{Kong2016}\cite{Ryzhov2023}\cite{Voltz1994}.

\textbf{Related work by the same authors.} In a previous work \cite{Nguyen2023}, we proposed to take into account the statistics of the AoA measurement error to improve the conventional reverse ray-tracing aided positioning technique. More specifically, it consists in launching the rays in a Monte Carlo manner, obtaining a map of points, and computing the position probabilities based on this map. In \cite{Nguyen2023}, the map of points is processed with the ``square method" described in Section~\ref{sec_process_map_of_points}. 
This paper proposes an additional improvement of this technique, regarding the treatment of the map of points, still considering the situation of noisy AoA measurements at the BS.

\textbf{Related work by other authors.} \cite{Kaya2007}\cite{Kong2006}\cite{Kong2016}\cite{Ryzhov2023}\cite{Voltz1994} consider a similar framework, using reverse ray tracing for positioning. In \cite{Kaya2007}\cite{Voltz1994}, no special treatment as a function of the AoA measurement noise is proposed (a treatment as a function of the signal power is described in \cite{Kaya2007}). In \cite{Kong2006}, the authors propose to launch several rays uniformly in an interval around the measured AoA. The considered width of the interval is
twice the estimated standard deviation of the estimated AoA value. In \cite{Kong2016}, the same authors propose to weight the obtained ray intersections (using the same launching method) with the ``dilution of precision" approach.
Note that \cite{Kong2006} is the benchmark used in the simulation results of \cite{Nguyen2023} (where a significant improvement is shown). The recent work \cite{Ryzhov2023} (see Sec. 4.2) contains similar ideas as our previous work \cite{Nguyen2023}: the statistics of the measurement error is considered in a Bayesian manner, by waiting the launched rays (and therefore the ray intersections) by the angle likelihood (similarly to Alg.~1 in \cite{Nguyen2023}). However, the more efficient Monte Carlo ray launching (Alg. 2 in \cite{Nguyen2023}) is not described.

The accuracy achieved by the proposed Monte-Carlo ray launching method with noisy measurements is similar (even slightly better) to the fingerprint methods based on the channel impulse response and neural networks. In \cite{TdocSumOct}\cite{Chatelier2023}, values around 1m are reported for the 90$\%$ quantile when neural networks are trained on dataset with inter-position spacing of 0.3m and no measurement error.

\textbf{Main contributions.}
For each BS, the Monte Carlo ray launching (introduced in \cite{Nguyen2023}) produces a map of points, obtained as the intersections of the rays with the xy plane.
 
As an improvement of the treatment of these maps of points, we propose to fit a parametric probability density function (pdf), such as a Gaussian mixture model (GMM), to each map.  
Then, the probability that the UE is in a position, given the measurement performed by each BS, is obtained by multiplying the fitted pdfs of each BS. This induces two main implementation modes:

\begin{itemize}
\item In the first mode, the rays are launched in the online phase and the pdfs are also computed in the online phase. The position probabilities are obtained by multiplying the pdfs. \\ \textbf{Advantage:} Fitting the pdfs enables to be more robust to a reduced number of launched rays.
\item The second mode is divided into an offline phase and an online phase. First, for each BS and for all possible AoA, the rays are launched to obtain the map of points. Parametric pdfs are fitted and their parameters are stored in a table. Then, in the online phase, an AoA is measured by each BS and the corresponding pdf parameters are recovered from the stored table. The pdfs are then used as in the first implementation mode. \\ \textbf{Advantage:} No ray launching in the online phase is required.
\end{itemize}

\section{Technical background}

Regarding the propagation of the rays, we consider only specular reflections. 
Indeed, due to smaller wavelength, the diffraction phenomenon is less important at millimeter-wave  \cite{Deng2016}\cite{Jacob2012} and terahertz bands \cite{Rap2019}. 
Consequently, this geometric assumption is made by many channel models as e.g., \cite{Lecci2020} and the above references on reverse ray tracing. Moreover, as the proposed algorithm do not utilize the signal power (only the geometric information), accurately simulating the signal-power attenuation in the propagation model is not critical.
Note that the considered model can be seen as a worst-case scenario from the positioning-ambiguity perspective: since we consider a relatively high number of reflected rays, including some rays that should not be detected with a more accurate propagation model, this induces many positioning solutions and thus many ambiguities.
However more realistic models, even at higher frequencies, should consider additional phenomena, see e.g., \cite{Muthineni2023} and also \cite{Bhatia2023A}\cite{Bhatia2023B} to understand how one can tune realistic ray-tracing models. 

Note also that the proposed approach could be extended to account for diffractions (e.g., for lower-frequency scenarios): At a diffraction point, the signal propagates in all directions, meaning that there is an uncertainty on the angle of the subsequent ray (in the reverse propagation). One should therefore generate several rays where the sum of the probabilities associated to these rays is equal to the probability of the generating ray. This can be done via probabilistic ray weighting \cite{Nguyen2023} or via Monte-Carlo. If needed, diffuse scattering on non-flat surfaces could also be handled in a similar way.

\subsection{Reverse ray-tracing positioning}

We first begin by recalling the reverse ray-tracing positioning method. For the sake of simplicity, we assume knowledge of the elevation of the UE to locate. This corresponds e.g., to an industrial scenario with moving robots on the floor.
Note that the method can be extended to a 3D positioning problem.

Given a measured AoA at one BS, ray launching from the BS in this direction is performed, i.e., the measured AoA becomes the angle of departure (AoD) of the rays. The process is repeated for all BS and the intersection of the rays is the estimated position, see Figure~\ref{fig:example_triang_reverse_ray}. 

\subsubsection{Monte-Carlo ray launching (proposed in \cite{Nguyen2023})} 
\begin{figure}[t]
\vspace{-15mm}
    \centering
    \includegraphics[scale=0.59]{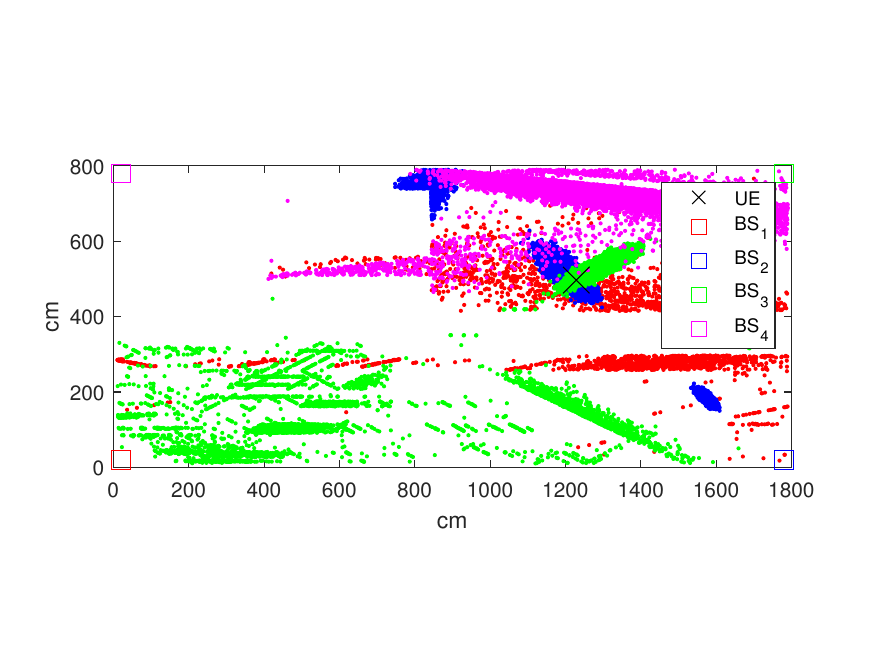}
\vspace{-14mm}
    \caption{Map of points corresponding to the positions where the rays launched by each BS cross the xy plane (having the dimensions provided in Section~\ref{sec_sim_env}). }
    \label{fig:map_points}
\vspace{-4mm}
\end{figure}

If the AoA measurements are noisy, we can proceed as follows: For each BS, the AoD of the rays to be launched are sampled according to the statistics of the AoA measurement error. Many rays are launched per BS. We refer to this method as the Monte Carlo approach.  The result of this ray-launching step is a set of points in the xy plane at the UE elevation. These points correspond to the position where the launched rays cross the xy plane. We call this set of points a map. An example of such a map is shown on Figure~\ref{fig:map_points} (using the simulation environment presented in Section~\ref{sec_sim_env}), with one color for the set of points corresponding to each BS. 

\subsubsection{Processing the map of points in \cite{Nguyen2023}}
\label{sec_process_map_of_points}
The position probabilities is then computed as follows. The map of points is first divided in set of squares. Within each square, one checks if there is at least one point associated to each BS. If this is the case, one counts the number of such set of points. The square containing the highest number is the most likely position. See Alg.~1 in \cite{Nguyen2023}. We refer to this approach as the ``square method".

With this latter method, the size of the squares is an important hyper-parameter: If the size is large then one loses accuracy (the position estimate is the center of the square). If the size of the squares is small, then there may not be a square gathering at least one point associated with each BS, which results in a failure. As a result, the size is chosen as the smallest such that the number of failures remains low (accuracy/robustness trade-off). Unsurprisingly, the higher the number of launched rays, the smaller the size can be.
Hence, the smaller the number of launched rays, the lower the accuracy, see Figure~4 in \cite{Nguyen2023} (and also Figure~\ref{fig:perf_algo} at the end of the paper). 

\subsection{Statistical modeling and notations}

Let $X$ be a random variable representing the position of a UE. Let $\Theta$ be a random variable representing the true AoA of the signal, and $Y$ a random variable representing the BS measurement(s) of the AoA.  The set of possible values $\theta$ for $\Theta$ is denoted by $\mathcal{C}$. For instance, $\mathcal{C}=[0,2\pi[$ in the 2D case or $[0,\ 2\pi [ \times [0,\ \pi[$ in the 3D case. 
 We let $n$ be the number of measurements such that $\textbf{y} = [y_1,y_2,\ldots,y_i,\ldots,y_n]$.  
The vector $\textbf{y}$ comprises the measurements performed by all BS. For the sake of simplicity, we assume that there is one measurement per BS, and therefore $n$ BS. Regarding the notations, we use $p\left(\theta\middle| y\right)$ for $p\left(\Theta=\theta|Y=y\right)$ and similarly $p\left(x\middle| y\right)$ for $p\left(X=x|Y=y\right)$. The distribution $p\left(\theta\middle| y\right)$ represents the statistics of the uplink AoA measurement error, e.g., $p\left(\theta\middle| y\right) \propto p\left(y \middle| \theta \right) \sim \mathcal{N}(\theta , \sigma^2)$. \\
In the considered problem, the goal is to compute  $p\left(x|\textbf{y}\right)$.

\section{Probability density function fitting}

In order to address the accuracy and robustness issues encountered with a low number of rays, we propose to fit a 2D parametric pdf on the map of points obtained when launching the rays from one BS in a Monte Carlo manner. One pdf is fitted to the map obtained by each BS. Using a continuous 
distribution enables to have a probability value for each position of scene.

\subsection{Formulation of the problem}

We define $p\left(x\middle| y_i\right)$ as the probability distribution that the uplink signal yielding the measure $y_i$ was transmitted from the position $x$. The result of the fitting step is thus, for each BS $i$, a distribution $p\left( x  \middle| y_i\right)$.
Recall that the statistics of the AoA $\theta_i$ at one BS is characterized by the distribution $p(\theta_i|y_i)$. The distribution $p(x|y_i)$ can be expressed as a function of $p(\theta_i|y_i)$ by marginalization over $\theta_i$ as follows

\vspace{-4mm}
\begin{align}
\label{equ_distri_posi}
\vspace{-4mm}
\begin{split}
p\left(x \middle| y_i\right)&=\int_{\theta_i}{p\left(x\middle|\theta_i\right)p\left(\theta_i\middle| y_i\right)}d\theta_i. 
\end{split}
\vspace{-2mm}
\end{align}

If the rays cross only once the xy plane, the term $p\left(x\middle|\theta_i\right)$ is an indicator function (computed thanks to the digital twin): it is equal to 1 if the ray with AoD $\theta$ crosses $x$ and 0 otherwise. Hence, the above equation indicates (via the integral) that if one location $x$ is crossed by several rays, the probabilities $p\left(\theta_i \middle| y_i\right)$ corresponding to each ray should simply be added.


If one ray crosses several times the xy plane at two positions $x_1$ and $x_2$, then $p\left(x_1\middle|\theta_i \right)= p\left(x_2\middle|\theta_i \right)=0.5$, and $p\left(x_k\middle|\theta_i \right)=0$ for all other $x_k$. 
In general, the probability $p\left(x_k\middle|\theta_i \right)$ of the crossing locations $x_k$ are therefore $p\left(x_k\middle|\theta_i \right)=$  1$/$number of crossing location of the ray. 

As mentioned at the beginning of the section, the goal is to fit a distribution $p\left(x \middle| y_i\right)$ to the set of points obtained when launching the rays in a Monte Carlo manner. 
We recall that the Monte Carlo approach consists in implementing the marginalization problem \eqref{equ_distri_posi} by sampling the angles $\theta_i$  of the launched rays according to $p(\theta_i|y_i)$. Hence, \eqref{equ_distri_posi} becomes: $p\left(x \middle| y_i\right) \approx \sum_{\text{Rays with AoD  } \theta_i }\ p\left(x\middle|\theta_i\right)$. 
It means that the probability $p\left(x\middle| y_i\right)$ can be estimated by counting the number of rays crossing a given position $x$, i.e., the number of points at position $x$ on the map (and applying the weighting factor $p\left(x \middle|\theta_i \right)$ if a ray crosses several times the xy plane). As explained in Section~\ref{sec_process_map_of_points}, this operation is performed in \cite{Nguyen2023} by dividing the plane in small squares and counting the rays in each square. 
Alternatively, $p\left(x \middle| y_i\right)$ can be estimated via the fitting of a continuous parametric distribution.

\subsection{Choice of the continuous parametric distribution}

On Figure~\ref{fig:map_points}, we observe that the points obtained for each BS (one color) are grouped in clusters. 
The fitting problem is therefore similar to a clustering problem. 
Of course, the distribution of the points depends on the AoA measurement-error distribution.
Hence, the fitting algorithm should: 1- Find the clusters. 2- Fit a distribution to each cluster. 3- Assign weights to the clusters (as a function of the more likely clusters).
The ``default" distribution for each cluster can be a Gaussian distribution\footnote{Note however that there is no guarantee that it is the most suited distribution. Considering other distributions for the clusters is left for future work (even though, as shown in the simulation results, the gain with respect to the square method is already satisfying).}. 
As a result, this approach leads naturally to the choice of the GMM as distribution.
Note also that this distribution is parametrized by a relatively low number of parameters, which will be useful for the ``second implementation mode" described below.

\begin{figure}[t]
\vspace{-3mm}
    \centering
    \includegraphics[scale=0.59]{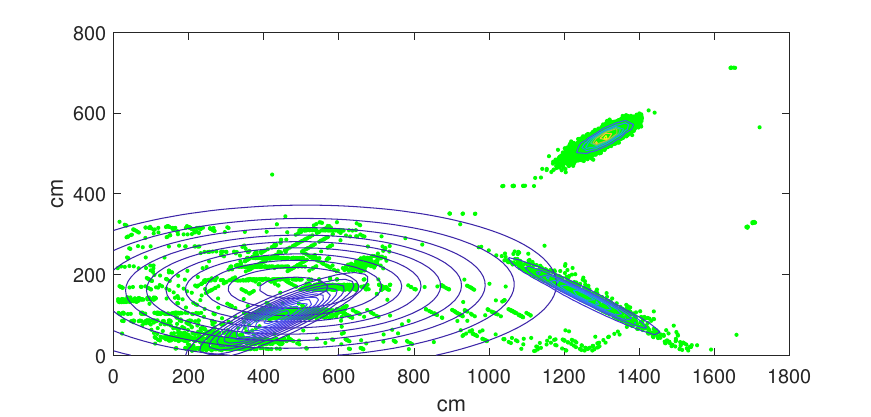}
\vspace{-2mm}
    \caption{Green: Map of points obtained when launching the rays from one BS with a given AoA and a given error statistics. Blue: Contour plot of the clusters of the fitted GMM.}
    \label{fig:fitted_GMM}
\vspace{-5mm}
\end{figure}

The standard approach to fit a GMM on a map of points is to use the expectation-maximization (EM) algorithm (see e.g., \cite{Roche2011}).
The main steps of the fitting algorithm are similar to the above bullet points.
Regarding the choice of the number of clusters, we run the algorithm for several values and we keep the model which yields the best Akaike information criterion \cite{Akaike1974}.
Figure~\ref{fig:fitted_GMM} shows an example of a fitted GMM to a map of points (same green points as on Figure~\ref{fig:map_points}) with the simulation environment described in Section~\ref{sec_sim_env}.

This process of fitting a pdf is repeated for each BS. As a result, we obtain $n$ distributions $p\left(x\middle| y_i\right)$, one for each measure $y_i$ performed at each of the $n$ BS.

\section{Exploiting the fitted probability density functions for positioning}

We now focus on the distribution $p(x|\mathbf{y})$, thus taking into account all measurements collected in the vector $\mathbf{y}$.
We show how $p(x|\mathbf{y})$ is expressed as a function of the distributions $p(x|y_i)$.

As proposed in \cite{Nguyen2023}, one can marginalize with respect to the angles $\theta$  as follows:
\vspace{-2mm}
\begin{align}
\begin{split}
p(x|\mathbf{y})
&=\int_{\boldsymbol{\theta} \in\mathcal{C}^n } p\left(x  \middle|\boldsymbol{\theta} \right)\prod_{i=1}^{n} p\left(\theta_i\middle| y_i\right)d\boldsymbol{\theta}, \\
&\propto \int_{\boldsymbol{\theta}\in\mathcal{C}^n} \prod_{i=1}^{n} p\left(x\middle|\theta_i\right)p\left(\theta_i\middle| y_i\right)d\boldsymbol{\theta}  = \prod_{i=1}^{n}  p(x|y_i),
\end{split}
\end{align}
where a uniformly-distributed prior $p(x)$ is thus assumed.

Alternatively, the same result can be obtained as:
\begin{align}
\begin{split}
&p\left(x \middle| \textbf{y} \right) =  \frac{p(\textbf{y}|x) p(x)}{p(\textbf{y})} = p(x) \prod_{i=1}^{n}  \frac{p(y_i|x)}{p(y_i)} \propto  \prod_{i=1}^{n}  p(x|y_i).
\end{split}
\end{align}

As a result, having the distributions $p\left(x\middle|y_i\right)$, for all $y_i$, one can directly compute $p\left(x\middle|\mathbf{y}\right)$ by multiplying the pdfs. This induces the two main implementation modes described below.

In the first implementation mode, ray launching and pdf fitting are performed for each new measurement set.
In the second implementation mode,  ray launching and pdf fitting are performed only once offline and the resulting distribution parameters  stored in a table. In the online phase, the suited parameters are recovered from the table based on the measurements.

\subsection{First implementation mode}

In this first implementation mode, the position estimation is realized as follows.
The AoA measurement error $\sigma^2$ is estimated offline by each BS. Then, in the online phase:

\begin{enumerate}
\item	Each BS performs an uplink AoA measurement $y_i$.
\item	Each BS launches the rays, in a Monte Carlo manner, with AoD angles sampled according to $p\left(\theta_i\middle| y_i\right)$.
\item	For each BS, the map of points where the rays cross the xy plane is constructed.   
\item	For each map, a parametric pdf for $p\left(x \middle| y_i\right)$, such as the GMM,  is computed.
\item	Finally, the distribution $p\left(x \middle|\mathbf{y}\right)$ is computed by multiplying the computed pdfs of all BS.	The maximum value of the distribution gives the most likely estimate.
\end{enumerate}

The simulation results in Section~\ref{sec_simu} show that this method is robust to a reduced number of launched rays.

\subsection{Second implementation mode}

While the first implementation mode is robust to a reduced number of rays, it still involves ray launching in the online phase, as well as fitting pdfs. This induces a high computational complexity and potential latency, discussed in Section~\ref{compu_complex_RT}.

This issue can be addressed by the following second implementation mode. The main idea is to compute and store the pdfs for each possible measured AoA $y_i$ in an offline phase. Then, in the online phase, each BS measures an angle $y_i$ and the corresponding stored distribution $p\left(x\middle| y_i\right)$ is recovered. Step 5 of the first implementation mode is then executed. The process is summarized below. 

\noindent \textbf{Offline phase:}  The AoA measurement error $\sigma^2$ is estimated offline by each BS. Then,
\begin{enumerate}
\item	Discretize the possible 3D angles $[0,2\pi[\times[0,\pi[$.
\item	For each BS, and for each discretized angle $y_i$, launch the rays in a Monte Carlo manner with AoD angles sampled according to $p\left(\theta_i\middle| y_i\right)$ and obtain the map of points where the rays cross the xy plane. 
\item	For each obtained map of points, fit a distribution $p\left(x\middle| y_i\right)$, such as a GMM, and store the parameters of the distribution. 
\end{enumerate}

\noindent \textbf{Online phase:} 
\begin{enumerate}
\item	An uplink AoA measurement $y_i$ is performed by each BS and the parameters of the distribution $p\left(x\middle| y_i\right)$ are recovered from the stored table.
\item	(=Step~5 of the first implementation mode) The distribution $p\left(x\middle|\mathbf{y}\right)$ is computed by multiplying the pdfs of all BS. The maximum value of the distribution gives the most likely estimate.
\end{enumerate}

As mentioned above, the main advantage of this second implementation mode is that it avoids ray launching and pdf fitting in the online phase.
Nevertheless, it requires storing the pdf parameters for each possible angle and each possible BS. 
Complexity estimates are provided in the following subsection.

\subsection{Complexity analysis}

\subsubsection{Complexity of ray tracing}
\label{compu_complex_RT}
The standard approach to implement a ray tracer is the following.
First, the reflecting surfaces of the environment are divided into triangles. 
For example a cubic room yields 12 triangles. As a result, an empty cubic scene involves a low number of triangles $t$. 
However, any object adds several triangles.
If it has not a well-defined geometrical shape, a piecewise linear approximation can be made. 
Hence, if the scene contains many objects, the number of triangles increases significantly. 
For example, the scene of Figure~\ref{fig:scene_digi-twin} involves 2164 triangles. A more realistic scene, with non-flat surface objects, would involve significantly more triangles. 

Then, once a ray is launched, one checks if there is an intersection with a triangle (low-complexity operation, with a constant number of flops $c$ for each check, see e.g., \cite{Vezina2024}).
Hence, the worst-case complexity of this step is the number of triangles multiplied by the constant $c$.
This operation has to be repeated the number of times the ray is reflected, determined by a maximum number of bounces $b$ (e.g., 5).

As a result, the complexity is $O(n \times t \times b)$, where $n$ is the number of launched rays.

Note that these operations can be easily parallelized as the triangle checks are independent. Moreover, the processing of each ray can also be done independently.
Therefore, the time complexity can be lower if the algorithm is implemented on GPU. However, this does not reduce the raw number of flops, which can thus be high if the scene is complex or the number of rays high. 

\subsubsection{Complexity of the square method of \cite{Nguyen2023} and GMM fitting}
The complexity of the square method is  $O(n + s )$, where $s$ is the number of squares (see Section~\ref{sec_process_map_of_points}), and with the assumption the the number of points is similar to the number of launched rays.

The EM algorithm for GMM fitting consists in alternating the E and M steps.
Both E and M steps involve performing operations on each data point for each cluster $k$. 
Hence, the complexity is $O(n\times k \times I)$, where $I$ is the number of iterations.

The remaining operations (e.g., multiplying the pdfs) have negligible complexity.
\subsubsection{Comparison of the computational complexity}
Based on the above discussion, we find the following estimated complexity (number of flops):
\begin{itemize}
\item Algorithm of \cite{Nguyen2023}: $O(n \times ( t \times b + 1) + s )$ $\approx O(n \times t \times b)$. 
\item First implementation mode : $O(n' \times ( t \times b + k \times I) )$ $\approx O(n' \times  t \times b)$, where $n'$ is the required number of rays to be launched to achieve the same performance as the square method with a high number of rays. In the reported simulations $n/n'=100$. 
\item Second implementation mode (online) : $O(1)$ (which corresponds to reading values in a table, where the index can be found in time independent of the table size).
\item Second implementation mode (offline) : $O(|\mathcal{C}| \times n' \times ( t \times b + k \times I) )$ (see the following subsection for the notation~$\mathcal{C}$).
\end{itemize}

Hence, the gain of the first implementation mode over \cite{Nguyen2023} comes mainly from the difference between $n$ and $n'$. 
The complexity of the second implementation mode (online) is independent from the parameter values.

\subsubsection{Storage complexity}

Let $N$ be the number of BS. In the simulation examples of Section~\ref{sec_simu}, we have $N=4$. Let $\left|\mathcal{C}\right|$ be the number of values used to discretized the possible angles. Let $P$ be the number of parameters of the fitted pdf. For instance, if a 2D GMM is used with 7 clusters, then $P=6\times 7=42$ parameters are required (where 6 represents the 2D mean value and the four values of the covariance). Then, the number of parameters to store is $ \left|\mathcal{C}\right| \times N \times P $.

Note that this is much smaller than the classical fingerprinting approaches where the number of parameters would behave either as $|\mathcal{C}|^N$ or as the number of discretized positions in the scene.

\vspace{-2mm}
\section{Simulation results}
\label{sec_simu}

\subsection{Simulation environment}
\label{sec_sim_env}

\begin{figure}[t]
    \centering
\vspace{-3mm}
    \includegraphics[scale=0.37, , angle =5 ]{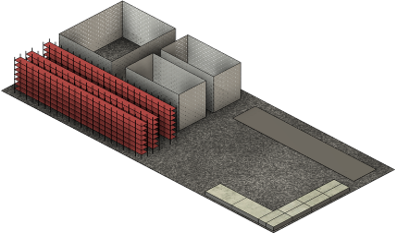}
\vspace{-3mm}
    \caption{Considered scene for the simulations.}
    \label{fig:scene_digi-twin}
\vspace{-4mm}
\end{figure}

The same numerical environment as the one described in \cite{Nguyen2023} is used. 
The scene is illustrated in Figure~\ref{fig:scene_digi-twin} and has the following dimensions: width 8m, length 18m, and height 2.5m. 
This environment is inspired from the recommendations of the 5G Alliance for Connected Industries and Automation (5G-ACIA) for indoor industrial scenario \cite{5G_ACIA}.
The 4 BS are located in the top corners of the scene.



\vspace{-4mm}
\subsection{Results}

The positioning error is denoted by $\mathcal{L} = ||\hat{x}_k - x_k||$, where $x_k$ is the true position and $\hat{x}_k$ the estimated position. 
The UE is randomly dropped in the scene (with a uniform distribution). Hence, there is randomness both in the UE position and the BS measurement errors.
A standard Gaussian AoA mesurement error model is used where $Y= \Theta +W$ with $W\sim \mathcal{N}(0,\sigma^2)$. The results are shown for several values of $\sigma^2$, namely 1 and 0.25. These values are chosen to show how large the measurement error can be while having a positioning error largely below 1 meter. We note that in some systems the AoA measurement error can be significantly smaller, e.g. the root mean square error can be as low as 0.01 degree in \cite{Wei2019}.

\begin{figure}[t]
    \centering
\vspace{-1mm}
\begin{subfigure}[b]{.24\textwidth}
    \includegraphics[scale=0.59]{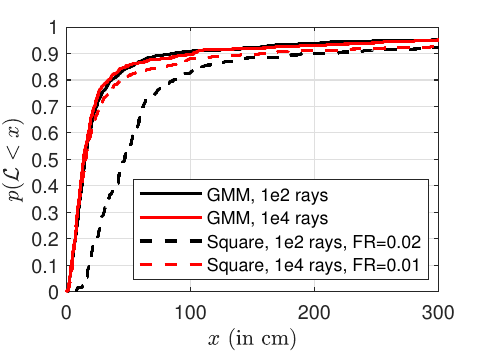}
    \caption{$\sigma^2=1$. }
\end{subfigure}%
\hfill
\vspace{-1mm}
\begin{subfigure}[b]{.24\textwidth}
    \centering
    \includegraphics[scale=0.56]{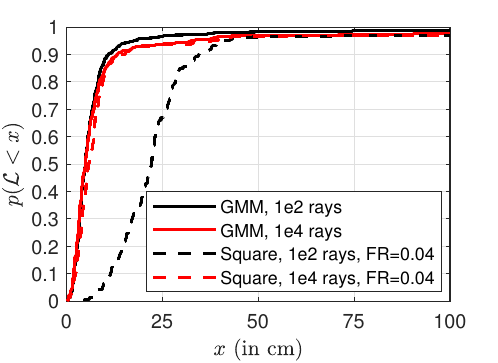}
    \caption{$\sigma^2=0.25$. }
\end{subfigure}%
    \caption{Cumulative density function of the positioning error. }
   \label{fig:perf_algo} 
\vspace{-4mm}
\end{figure}

Figure~\ref{fig:perf_algo} show the positioning accuracy obtained both with a low number of launched rays  per BS (100) and a high number of rays per BS (10000) using the GMM fitting approach for a measurement error $\sigma^2=1$ and $\sigma^2=0.25$.
The accuracy with the square method of \cite{Nguyen2023} is also shown on the figures\footnote{Note that we managed to slightly improve the performance of the benchmark square method with a limited number of rays (compared to e.g., Figure~4 in \cite{Nguyen2023}) by adding some additional processing also considering subsets of the 4 BS. The performance however remains significantly degraded.}. The notation FR means failure rate (for the square method), which means that we do not consider the UE positions not having at least one square with the rays of at least 3 out of the 4 BS. There is no such failure event in the case of the GMM approach.

One can see that the performance with the proposed method having a low number of rays is as good as the one having the high number of rays. As expected, the pdf fitting approach enables to maintain the performance with a reduced number of rays.
%
%

On Figure~\ref{fig:perf_algo} (b), we observe that the performance with a low number of rays (using the GMM approach) is even slightly better than the one with a high number of rays. This can be explained as follows. In our implemented algorithm, if the covariance matrix of one of the four fitted GMM models (one for each BS) is ill-conditioned, which happens more frequently with a low number of rays, then we use only the three remaining models. 
As a result, it might be possible to improve the algorithm by considering rays only from subsets of the BS even if the model matrices are not ill-conditioned. This is left for future work.

\vspace{-2mm}
\section{Conclusions}

In this paper, we proposed to improve a positioning technique based on noisy uplink AoA measurements and a digital twin of the environment. The main idea consists in fitting a parametric pdf, such as the GMM, on each map of points obtained via ray tracing from each BS. The position probabilities are then obtained by multiplying the fitted pdfs.
This solution contains two main advantages. First, unlike previous methods, the algorithm accuracy is robust to a reduced number of launched rays. Second, it enables to perform the ray launching operations in an offline phase, such that, in the online phase, the parameters of the pdfs are directly recovered from a table given the AoA measurements performed. This offers a significant complexity reduction.

\end{document}